\documentclass[runningheads]{llncs}
\usepackage[T1]{fontenc}
\usepackage{graphicx}
\begin{document}
\title{GenAI-Enabled Backlog Grooming in Agile Software Projects: An Empirical Study}
%
%
\author{Kasper Lien Oftebro \inst{1} \and Anh Nguyen-Duc\inst{1,2}\orcidID{0000-0002-7063-9200} \and Kai-Kristian Kemell \inst{3} 
} 
\authorrunning{Oftebro et al.}
\institute{
Norwegian University of Science and Technology, Trondheim, Norway 
\and
University of South-Eastern Norway, Bø i Telemark, Norway 
\and
Tampere University, Tampere, Finland 
\\
\email{kasper.oftebro@gmail.com, angu@usn.no, kai-kristian.kemell@tuni.fi}
}

\maketitle              
\begin{abstract}
Effective backlog management is critical for ensuring that development teams remain aligned with evolving requirements and stakeholder expectations. However,
as product backlogs consistently grow in scale and complexity, they tend to become cluttered with redundant, outdated, or poorly defined tasks, complicating prioritization and decision-making processes. This study investigates whether a generative-AI (GenAI) assistant can automate backlog grooming in Agile software projects without sacrificing accuracy or transparency. Through Design Science cicles, we developed a Jira plug-in that embeds backlog issues with the vector database, detects duplicates via cosine similarity, and leverage GPT-4o model to propose merges, deletions, or new issues. We found that AI-assisted backlog grooming achieved 100\% precision while reducing the time-to-completion by 45\%. The findings demonstrated the tool’s potential to streamline backlog refinement processes while improving user experiences.

\keywords{Automated grooming  \and Large language model \and Agile software management \and Product Backlog \and GenAI for project management}
\end{abstract}
\section{Introduction}
The rapid advancement of Generative Artificial Intelligence (GenAI), particularly large language models (LLMs), is opening new possibilities for automating complex, knowledge-intensive tasks in software engineering. While much attention has been placed on how GenAI supports software developers—such as through code generation, documentation, and testing \cite{he_llm-based_2024,wang_review_2023,hou_large_2024}, its application in software project management remains underexplored \cite{nguyenduc2024chatgpt}. For project managers and agile team leads, managing evolving priorities, stakeholder needs, and task complexity is critical to project success. Here, GenAI offers the potential to assist in maintaining strategic oversight, refining priorities, and streamlining coordination, areas that have not received sufficient research focus compared to technical development support.

Agile software development is a dominant and impactful methodology in modern IT projects, characterized by its iterative processes, customer collaboration, and responsiveness to change \cite{dyba_empirical_2008,dingsoyr_decade_2012}. At the heart of Agile practice lies the product backlog—a dynamic, evolving list of features, issues, and improvements. Maintaining an effective product backlog is essential but increasingly difficult as projects scale. Overtime, backlogs can become cluttered with redundant, outdated, or poorly defined items, making it difficult
for teams to efficiently deliver value \cite{ramesh2010agile}. The dynamic nature of software projects often leads
to overlooked requirements, further complicating backlog refinement and prioritization \cite{bjarnason2011casestudy}.

Manual grooming is time-consuming and error-prone, requiring teams to spot duplicate items, outdated issues, and misaligned priorities without systematic support. Although tools like Jira provide organizational structure, they lack sufficiently intelligent automation to proactively suggest optimizations or flag redundant items. The result is often a cluttered backlog that hinders transparency, delays delivery, and burdens team morale. Advancements in GenAI have shown promise in automating complex tasks in software engineering, yet their application in backlog management remains under-explored. There is a gap that GenAI could fill by automating the identification of redundant or outdated tasks, facilitating the merging of related tasks into cohesive user stories, and suggesting missing items aligned with project objectives.

The primary objective of this research is to evaluate the effectiveness and accuracy of an AI-based solution for detecting and resolving redundant backlog items within the Jira management system. Our Reseach Question is:
\begin{itemize}
    \item How does GenAI-enabled automated backlog grooming perform compared to manual approaches in Agile software project management?
\end{itemize}

This research contributes to both academic understanding and industry practice. For research, the integration of GenAI into software project management, particularly in backlog management, represents a promising but underexplored topic. Existing research on AI-driven backlog refinement is sparse, with only a few frameworks leveraging natural language processing (NLP) and machine learning for automating these tasks. The proposed tool, informed by frameworks such as the "Smart AI Framework for Backlog Refinement" \cite{nasiri2024backlogrefinement}, aims to automate key aspects of backlog management, thereby reducing manual effort and improving task prioritization. This study builds on these foundations to apply generative AI more broadly in project management \cite{dossantos2024userstorygeneration}.  For practice, the study proposes a practical AI-integrated solution for backlog refinement, demonstrating how GenAI can reduce manual effort, improve clarity, and support more inclusive and accurate prioritization.

The study is organized as follows: Section 2 presents related work in the application of GenAI in Software Engineering. Section 3 presents the development of our grooming approach. Section 4 presents our experimental setting. Section 5 is our result. Section 6 and Section 7 discusses and concludes the paper.
\section{Related Work}
\subsection{Applications of GenAI in Software Engineering}
GenAI is currently improving efficiency, code quality, and project management within software engineering. Recent research shows that GenAI methods, when implemented thoughtfully, hold the potential to improve quality and efficiency in software development \cite{he_llm-based_2024,hou_large_2024,wang_review_2023}. For instance, Peng et al. reported that developers using GitHub Copilot completed a standardized programming task 55.8\% faster than those without AI assistance \cite{peng2023_23}. Calegario et al. emphasize that successful AI adoption requires thoughtful implementation, but when done right, AI can yield substantial improvements in productivity and code quality through task automation, intelligent suggestions, and optimization \cite{calegario2023_22}.

Beyond coding, AI is increasingly being used to enhance project management by automating decision-making, forecasting timelines, identifying risks, and allocating resources more efficiently \cite{crawford2023_24}. Strategic implementation remains crucial for realizing these benefits \cite{Vayyavur2024_25}. In particular, the integration of Natural Language Processing (NLP) into backlog management has shown promise. NLP techniques can classify, clarify, and refine user stories or requirements, helping engineers reduce ambiguity and redundancy in backlogs \cite{liu2024ai_26}. 

\subsection{Vector embeddings for Backlog item similarity}
Vector embeddings is a Natural Language Processing (NLP) approach that encodes textual data into dense numerical representations that capture semantic relationships. Unlike traditional techniques such as Term Frequency-Inverse Document Frequency (TF-IDF), which rely on word frequency statistics and struggle with understanding meaning, modern embedding methods like Word2Vec, BERT, and Sentence Transformers incorporate context, syntax, and word order to deliver richer representations of meaning. Shahmirzadi et al. have demonstrated that while TF-IDF is suitable for basic keyword overlap detection, neural embeddings significantly outperform it when identifying deeper semantic similarities in real-world data \cite{shahmirzadi2018textsimilarityvectorspace}.

In backlog management, embedding techniques can be applied in several ways. One of the most prominent use cases is duplicate detection, where vector representations of issues are compared using cosine similarity to identify items with high semantic overlap. This allows teams to consolidate redundant tasks and improve backlog clarity. Moreover, embedding-powered semantic search supports breaking down larger queries to discover related items that might be phrased differently but describe similar functionality. Beyond redundancy management, vector embeddings also support dependency resolution by mapping the relationships between tasks and identifying inconsistencies or gaps, as discussed by Raatikainen et al. \cite{siddique2022ai_requirements}. These applications position vector embeddings as a foundational tool for intelligent backlog refinement, enabling more scalable and context-aware Agile project management.
\section{Prototype development}

This study adopts a Design Science Research (DSR) methodology complemented by user testing sessions and empirical evaluation to investigate and enhance backlog management in software development projects \cite{hevner_design_2004}. DSR is chosen for its emphasis on the creation and evaluation of innovative artifacts designed to solve identified problems within a specific context. By leveraging DSR, this research aims to develop a GenAI prototype integrated with Jira, iteratively refining its functionality based on user feedback and empirical
data.
The intention of this tool is the following.
\begin{itemize}
    \item Detection of redundant issues: Use of vector embeddings to detect and present semantically similar issues in the backlog as potential duplicates.
    \item Merge related tasks: Suggest the consolidaton of similar issues into cohesive user stories to improve clarity and manageability.
\end{itemize}
 
\begin{figure}
    \centering
    \includegraphics[width=0.5\linewidth]{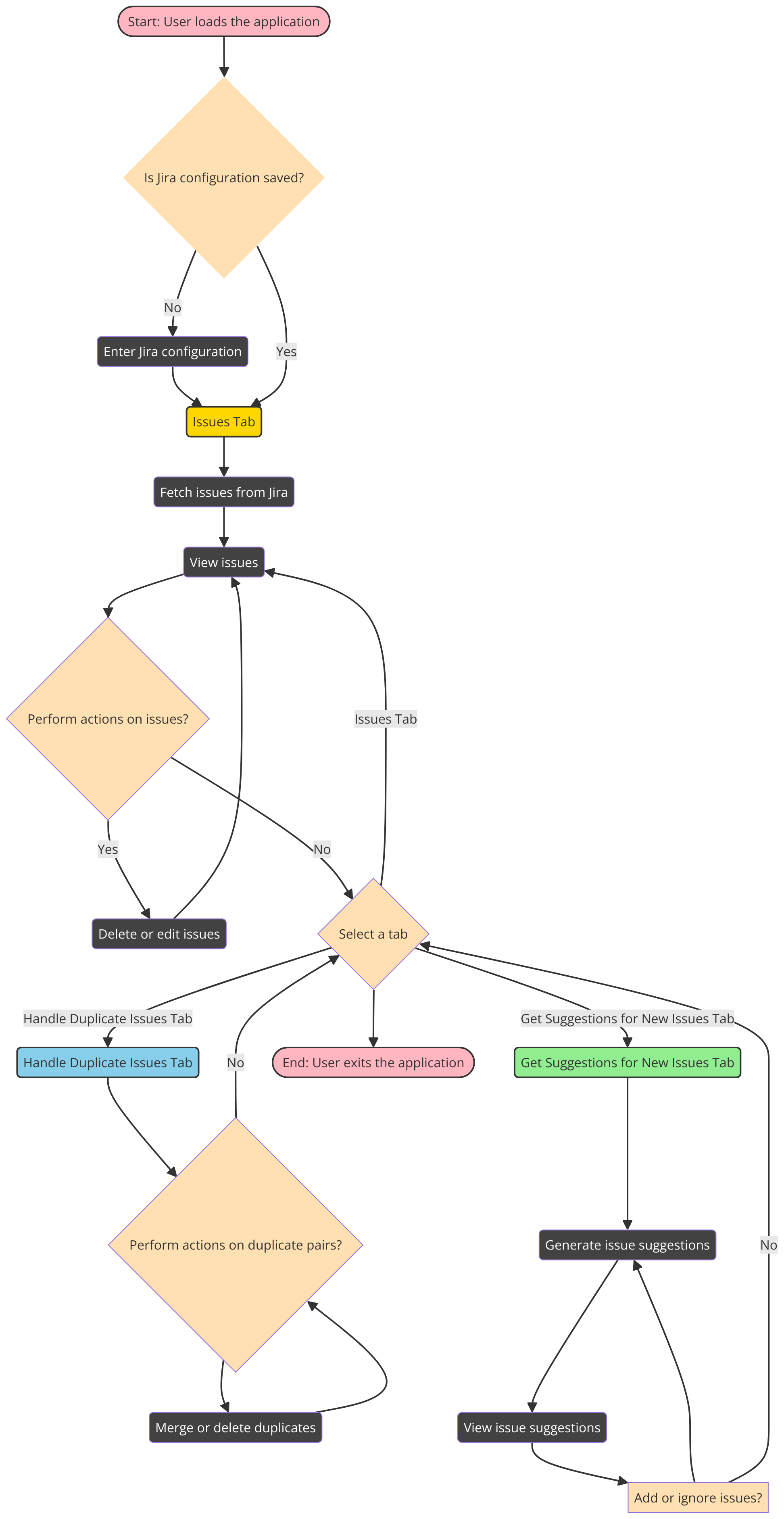}
    \caption{Workflow of the Generative AI Tool for Backlog Management}
    \label{fig:conceptual_framework}
\end{figure}

This prototype integrates several AI-driven modules into a cohesive system designed to enhance backlog management in Agile software projects. Its key components—semantic analysis, generative modeling, user interaction, and Jira integration—work together to automate backlog refinement tasks, all while keeping human oversight at the center. Figure \ref{fig:conceptual_framework} illustrates the overall workflow and how each module contributes to improving backlog clarity, completeness, and efficiency.

The duplicate detection module leverages semantic embeddings (via the text-embedding-3-large model) to compute cosine similarity scores among existing Jira items. Backlog entries that exceed a defined similarity threshold are flagged as potential duplicates. The model, supported by GPT-4o, then provides contextual recommendations for resolving these, such as merging related issues or eliminating obsolete ones. In parallel, the issue suggestion module applies generative AI techniques to propose new backlog items derived from the project description, current issues, and optional user prompts. Suggested tasks are filtered through similarity checks to avoid redundancy and are prioritized according to their alignment with project goals. The tool was extended to suggest new issues, leveraging generative AI to fill potential gaps in the backlog. These suggestions were derived from project descriptions, current backlog content, and optional user inputs, and were filtered for redundancy before being presented to the user.

All AI outputs are presented in a tabular user interface, designed to retain human control and foster trust. Users can inspect, accept, modify, or reject suggestions. No changes are applied without user confirmation. This hybrid human-AI interaction model ensures that automation supports, rather than replaces, the team’s judgment. Finally, real-time Jira integration allows the tool to synchronize accepted actions—merging duplicates, creating new issues, or updating statuses—directly into the Jira environment. This minimizes disruption to existing workflows and ensures project consistency across tools.

\section{Methodological approach}

\subsection{Experiment setting} 
Five user testing sessions were conducted within a controlled environment. Participants are senior project managers and test managers with experience varied from 10 to 20+ years within the domain of agile software projects. They are from two Nordic mid-size IT consulting companies. A Jira project consisting of 30 known issues. The tested tool, with the correct credentials so it is set up in accordance with the Jira project.
Participants were instructed to complete a set of predefined tasks, with no intervention:
\begin{enumerate}
    \item Fetch issues from the Jira backlog
    \item Use the tool to find duplicate backlog items
    \item Resolve three identified duplicates
\end{enumerate} 
Particularly, testers were gived distinct workflows:
\begin{itemize}
    \item Two testers (\#9, \#12) worked only with the tool, but always had the possibility to navigate to the backlog in Jira to validate changes or use both in combination in other ways. 
    \item One tester (\#10) performed only manual grooming directly in Jira, and was afterwards shown the tool for qualitative comparison
    \item Two testers (\#8, \#11) carried out manual grooming first, and then used the tool allowing for direct qualitative comparison on both performance and specifically satisfaction. 
\end{itemize}

\subsection{Evaluation metric}
\begin{figure}
    \centering
    \includegraphics[width=\linewidth]{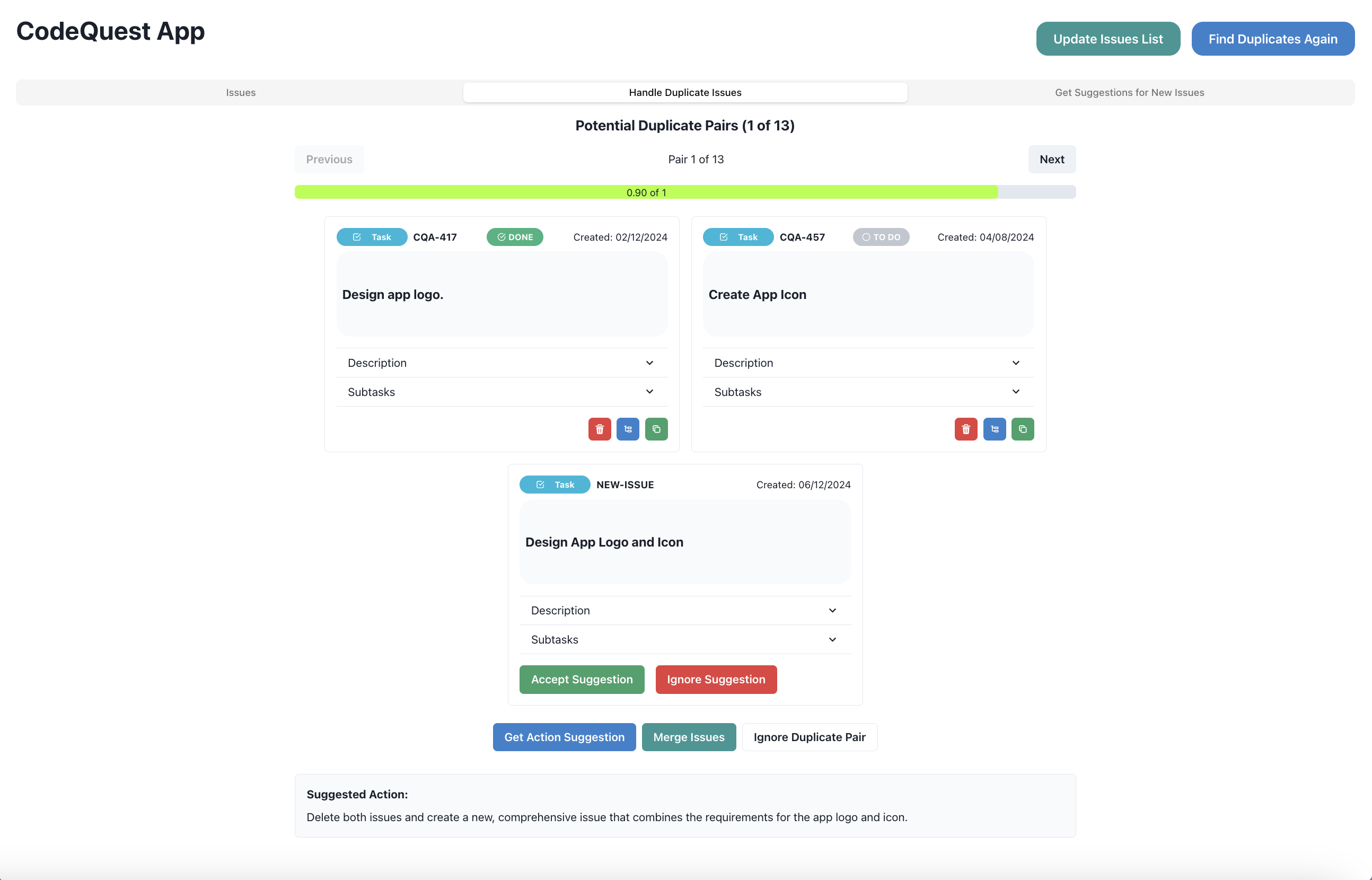}
    \label{fig:decision2}
    \caption{Final version of the Backlog Grooming Tool - decision-making interface}
    \label{fig:version2.0}
\end{figure}

The version tested can be seen in Figure \ref{fig:version2.0}.  Accuracy, precision, recall, and F1-scores were calculated to evaluate the effectiveness of both manual and GenAI-assisted backlog grooming performed by the testers. These metrics were based on the identification of duplicate Jira issues, with a ground truth established through a thorough manual review of the backlog to validate all true duplicate pairs. Precision measured the proportion of correctly identified duplicates among those flagged by the tool, while recall assessed the proportion of true duplicates successfully detected. The F1-score provided a balanced measure of the tool's performance, especially in cases where the number of duplicates varied significantly across projects. Accuracy reflected the overall correctness, including both duplicate and non-duplicate classifications.

In addition to performance metrics, time-to-completion data was collected to assess efficiency. This metric captured the total time each tester spent completing the backlog grooming task, with the endpoint determined by the user based on when they felt the work was sufficiently complete. However, to ensure comparability, time data was further analyzed against a reference point defined by an expert-reviewed, data-sensitive notion of task completeness. This allowed us to distinguish between subjective perceptions of task completion and objective closeness to a fully refined, duplication-free backlog.

\section{Results}
\subsubsection{Performance: }
\begin{figure}
    \centering
    \includegraphics[width=0.8\linewidth]{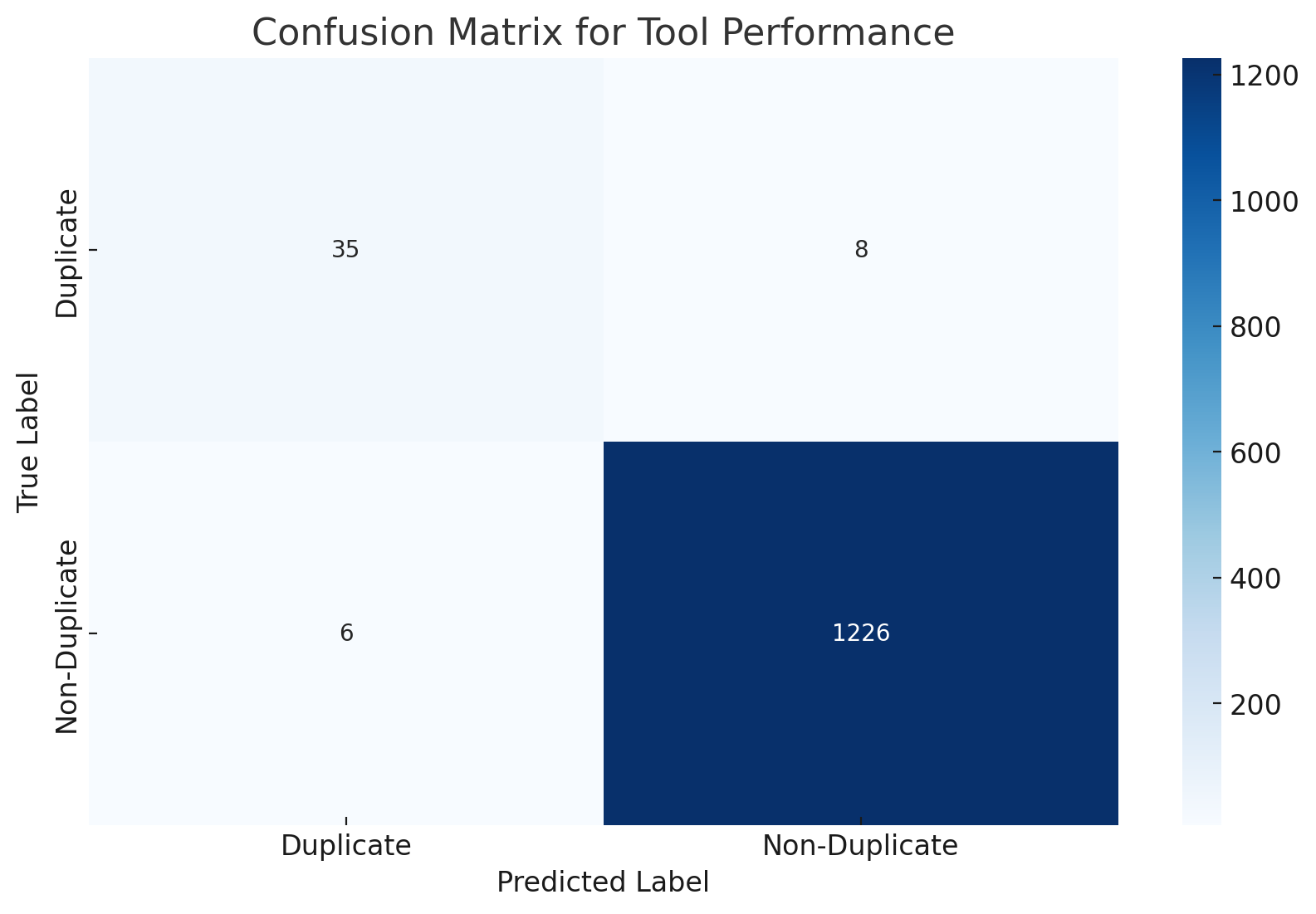}
    \caption{Confusion matrix showing the performance of the tool operating without human intervention, highlighting true and false predictions for duplicate and non-duplicate labels.}
    \label{fig:tool-only performance}
\end{figure}

The results focus on
the detection of duplicate issues in a labeled Jira backlog and show key metrics such as true positives (TP), false positives (FP), false negatives (FN) and true negatives (TN).
The labeled backlog contained a total of 41 true positives (issues that refer to the same functionality). The tool, when operated without human oversight (i.e. "auto" mode, where all suggestions are accepted), successfully identified 35 of these pairs as duplicates (TP), missed
6 pairs (FN), and introduced 8 FP, as seen in the confusion matrix in Figure \ref{fig:tool-only performance}. Notably, the FPs suggested by the tool were in every case rejected by the participants using the tool in interactive sessions, highlighting their critical thinking.

Table \ref{tab:metrics_results} presents the comparison between manual grooming done by our testers and automated grooming. Manual grooming alone provided significantly fewer detected duplicates. For instance, tester \#10, relying solely on manual methods, achieved only 7 TPs, while tester \#8 and Tester \#11 manually identified 8 and 12 TPs respectively. When the same testers made use of the tool, their TPs increased while their time to completion decreased. Tester \#8's TPs rose from 8 to 12, while their time to completion decreased from 25 to 11 minutes. Tester \#11 had an increase of 8 TPs, rising from 12 to 20, at the same time slicing their time to completion in half, from 24 minutes to 12 minutes. The final testers, \#9 and \#12, who only performed grooming using the tool, correctly found 21 and 12 TPs in 15 and 20 minutes respectively. 

\begin{table}[ht]
\centering
\caption{Comparing performance of manual grooming vs. automated grooming}
\begin{tabular}{|l|r|r|r|r|r|r|r|r|r|}
\hline
\textbf{Participant} & \textbf{TP} & \textbf{FP} & \textbf{FN} & \textbf{TN} & \textbf{Time (min)} & \textbf{Precision} & \textbf{Recall} & \textbf{Accuracy} & \textbf{F1} \\ \hline
\#8 Manual            & 8           & 1           & 33          & 1233        & 25                 & 0.8889            & 0.1951          & 0.9733           & 0.3200      \\ \hline
\#8 Auto              & 15          & 0           & 26          & 1234        & 11                 & 1.0000            & 0.3659          & 0.9796           & 0.5986      \\ \hline
\#11 Manual           & 12          & 1           & 29          & 1233        & 24                 & 0.9231            & 0.2927          & 0.9765           & 0.4444      \\ \hline
\#11 Auto             & 20          & 0           & 21          & 1234        & 12                 & 1.0000            & 0.4878          & 0.9835           & 0.6557      \\ \hline
\#9 Auto              & 21          & 0           & 20          & 1234        & 15                 & 1.0000            & 0.5122          & 0.9843           & 0.6774      \\ \hline
\#10 Manual           & 7           & 2           & 34          & 1232        & 25                 & 0.7778            & 0.1707          & 0.9718           & 0.2800      \\ \hline
\#12 Auto             & 12          & 0           & 29          & 1234        & 20                 & 1.0000            & 0.2927          & 0.9773           & 0.4528      \\ \hline
\end{tabular}
\label{tab:metrics_results}
\end{table}

\subsubsection{Efficiency: }
\begin{figure}
    \centering
    \includegraphics[width=0.8\linewidth]{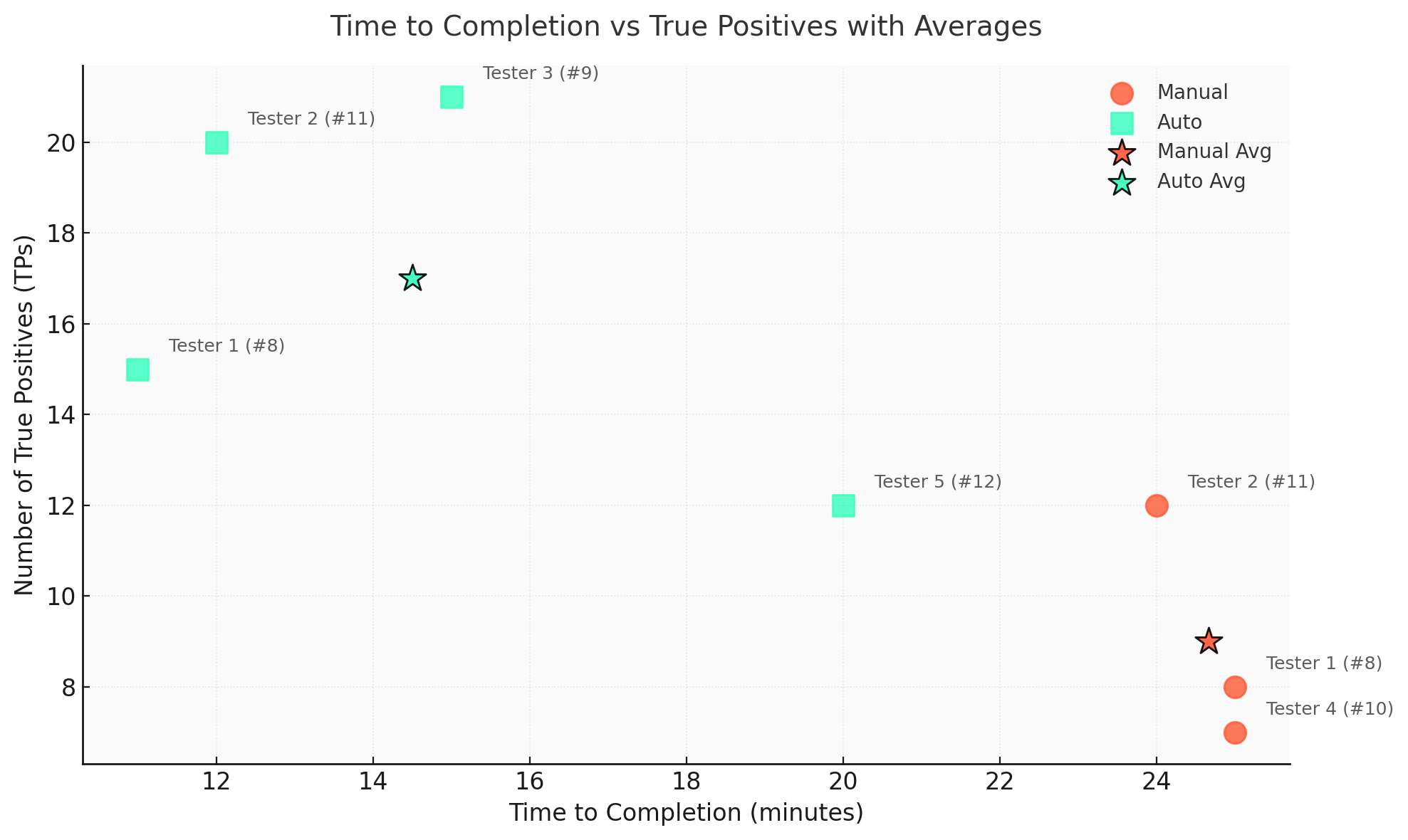}
    \caption{Time-to-completion versus the number of true positives detected across manual and automated grooming.}
    \label{fig:manual_vs_tool_assisted}
\end{figure}

When performing manual grooming, the participants all used 24-25 minutes before either opting out themselves or being interrupted by the test leader to manage time. The number of duplicates detected during these grooming sessions remained relatively low (ranging from 7 to 12 TPs). Using normalization techniques with regards to "completion degree" (i.e. how many duplicates found relative to the total number of 41), manual grooming displayed a much higher time-per-duplicate ratio with an average of 2 minutes and 54 seconds per detected duplicate, compared to 1 minute 35 seconds when using the tool as an assistant, showing a decrease of almost 50\% per duplicate pair (45.38\%). Comparisons on efficiency between tool-assisted performances and manual performances can be seen in figure \ref{fig:manual_vs_tool_assisted}

\subsubsection{User Experience:}
\begin{figure}
    \centering
    \includegraphics[width=0.8\linewidth]{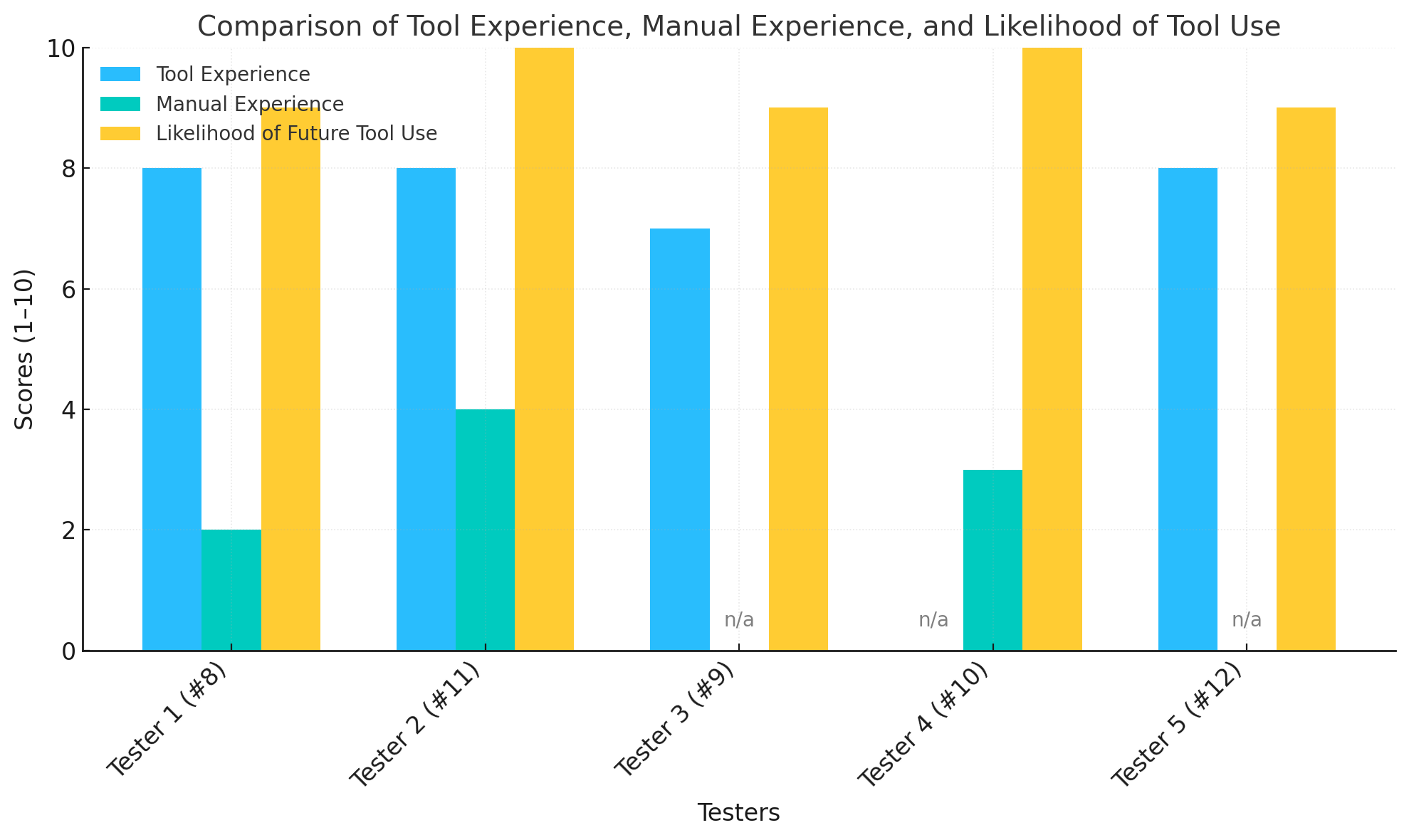}
    \caption{User experienced rated by Testers}
    \label{fig:qualitative_comparison}
\end{figure}

All testers were asked to rate their experience, boasting only scores ranging from 7-9 out of 10. By comparison, the user experience ratings of performing the manual grooming (\#8, \#10, and \#11) were all between 2-4, showing the positive impact the tool had on the overall user experience. The testers were also asked to rate their likelihood of adopting the tool in their workflows, resulting in only positive annotations. The received qualitative ratings can be seen in Figure \ref{fig:qualitative_comparison}

\section{Discussion and Conclusions}
\subsection{Implications of findings}
Nguyen-Duc and Khanna called for implementing GenAI in further tasks to reduce workload \cite{nguyenduc2024chatgpt}. This study provides quantitative evidence of workload reduction, demonstrating a decrease of approximately 45\% in time-to-completion for backlog grooming tasks. Precision rates reached 100\% for every participant using the tool to groom the backlog, while all other metrics improved as well. This highlighted the
tool’s ability to identify a broader range of relevant duplicates in a shorter time-span, while its implementation as an assistant rather than a decision-maker eliminated all false positives. 

The findings demonstrate how GenAI and vector embeddings have the ability to identify redundant backlog items and suggest actionable resolutions, underscoring its ability to streamline certain agile workflows. Through these promising efficiency gains, agile teams would be able to focus on higher-value activities such as strategic planning and stakeholder engagement. Despite the outlined potential, there are some key challenges in need of further review:
\begin{itemize}
    \item Domain-Specific Training: The performance metrics were captured from a labeled test dataset, made to resemble a small real-world project. This means that the application and potential effectiveness in diverse, domain-specific environments is implied, but unexplored. Training similar GenAI models with specific domain data is an interesting avenue for future research.
    \item User Trust and Transparency: The "black-box" nature of AI decision-making processes poses a potential barrier. Even though all users in this particular study seemed to look beyond it and were satisfied with the transparency and human-AI-collaboration aspect of the tool, that is not necessarily the case for everyone.
    \item Integration with Existing Systems: The developed system integrates with Jira, a preferred backlog management tool by many. Not everyone uses Jira though, and for it to integrate with other systems, there would be need for adjustments.
\end{itemize} 

\subsection{Limitation}
Despite the promising results in validation testing of the developed tool, several limitations
must be acknowledged. The tool’s performance was validated within a controlled testing environment a labeled dataset. Its applicability to a wider range of projects, potentially more domain specific, remains unexplored. Variations in terminologies and requirements documentation across different fields may impact effectiveness. Performance was tested on a relatively small backlog, and the computational demands
of larger datasets have not been tested. Ensuring the tool’s ability to handle extensive backlogs with precision and efficiency would be substantial.The sample size for both the validation testing was rel-
atively small. Although participants varied in role and experience, the limited number of testers may reduce the generalizability of the findings. A larger, more diverse participant group could provide stronger evidence regarding the tools performance

\section{Conclusions}
The research question— How does GenAI-enabled automated backlog grooming perform compared to manual approaches in Agile software project management? —is
conclusively addressed by our empirical findings. Specifically, employing embedding-based detection coupled with GPT-driven suggestions not only automated the identification of duplicates
with high precision but also streamlined backlog refinement to nearly half the time reported for manual methods. Consequently, these insights validate that GenAI and vector embeddings present a viable path for practical enhancements within IT project backlog
management.

\bibliographystyle{splncs04}
\bibliography{main.bib}

\begin{thebibliography}{10}
\providecommand{\url}[1]{\texttt{#1}}
\providecommand{\urlprefix}{URL }
\providecommand{\doi}[1]{https://doi.org/#1}

\bibitem{bjarnason2011casestudy}
Bjarnason, E., Wnuk, K., Regnell, B.: A case study on benefits and side-effects of agile practices in large-scale requirements engineering. In: Proceedings of the 1st Workshop on Agile Requirements Engineering. AREW '11, Association for Computing Machinery, New York, NY, USA (2011). \doi{10.1145/2068783.2068786}, \url{https://doi.org/10.1145/2068783.2068786}

\bibitem{calegario2023_22}
Calegario, F., Burégio, V., Erivaldo, F., Andrade, D.M.C., Felix, K., Barbosa, N., da~Silva~Lucena, P.L., França, C.: Exploring the intersection of generative ai and software development (2023), \url{https://arxiv.org/abs/2312.14262}

\bibitem{crawford2023_24}
Crawford, T., Duong, S., Fueston, R., Lawani, A., Owoade, S., Uzoka, A., Parizi, R.M., Yazdinejad, A.: Ai in software engineering: A survey on project management applications (2023), \url{https://arxiv.org/abs/2307.15224}

\bibitem{dingsoyr_decade_2012}
Dingsøyr, T., Nerur, S., Balijepally, V., Moe, N.B.: A decade of agile methodologies: {Towards} explaining agile software development. Journal of Systems and Software  \textbf{85}(6),  1213--1221 (Jun 2012). \doi{10.1016/j.jss.2012.02.033}, \url{https://www.sciencedirect.com/science/article/pii/S0164121212000532}, number: 6

\bibitem{dyba_empirical_2008}
Dybå, T., Dingsøyr, T.: Empirical studies of agile software development: {A} systematic review. Information and Software Technology  \textbf{50}(9),  833--859 (Aug 2008). \doi{10.1016/j.infsof.2008.01.006}, \url{https://www.sciencedirect.com/science/article/pii/S0950584908000256}

\bibitem{he_llm-based_2024}
He, J., Treude, C., Lo, D.: {LLM}-{Based} {Multi}-{Agent} {Systems} for {Software} {Engineering}: {Literature} {Review}, {Vision} and the {Road} {Ahead} (Dec 2024). \doi{10.48550/arXiv.2404.04834}, \url{http://arxiv.org/abs/2404.04834}, arXiv:2404.04834 [cs]

\bibitem{hevner_design_2004}
Hevner, A.R., March, S.T., Park, J., Ram, S.: Design {Science} in {Information} {Systems} {Research}. MIS Quarterly  \textbf{28}(1),  75--105 (2004). \doi{10.2307/25148625}, \url{https://www.jstor.org/stable/25148625}, publisher: Management Information Systems Research Center, University of Minnesota

\bibitem{hou_large_2024}
Hou, X., Zhao, Y., Liu, Y., Yang, Z., Wang, K., Li, L., Luo, X., Lo, D., Grundy, J., Wang, H.: Large {Language} {Models} for {Software} {Engineering}: {A} {Systematic} {Literature} {Review}. ACM Trans. Softw. Eng. Methodol.  \textbf{33}(8),  220:1--220:79 (Dec 2024). \doi{10.1145/3695988}, \url{https://dl.acm.org/doi/10.1145/3695988}

\bibitem{liu2024ai_26}
Liu, K., Reddivari, S., Reddivari, K.: Artificial intelligence in software requirements engineering: State-of-the-art. In: 2022 IEEE 23rd International Conference on Information Reuse and Integration for Data Science (IRI). pp. 106--111 (2022). \doi{10.1109/IRI54793.2022.00034}

\bibitem{nasiri2024backlogrefinement}
Nasiri, S., Lahmer, M.: A smart ai framework for backlog refinement and uml diagram generation. International Journal of Advanced Computer Science and Applications  \textbf{15} (01 2024). \doi{10.14569/IJACSA.2024.0150474}

\bibitem{nguyenduc2024chatgpt}
Nguyen~Duc, A., Khanna, D.: Value-Based Adoption of ChatGPT in Agile Software Development: A Survey Study of Nordic Software Experts, pp. 257--273 (06 2024). \doi{10.1007/978-3-031-55642-5_12}

\bibitem{peng2023_23}
Peng, S., Kalliamvakou, E., Cihon, P., Demirer, M.: The impact of ai on developer productivity: Evidence from github copilot (2023), \url{https://arxiv.org/abs/2302.06590}

\bibitem{ramesh2010agile}
Ramesh, B., Cao, L., Baskerville, R.: Agile requirements engineering practices and challenges: An empirical study. Inf. Syst. J.  \textbf{20},  449--480 (09 2010). \doi{10.1111/j.1365-2575.2007.00259.x}

\bibitem{dossantos2024userstorygeneration}
dos Santos, C., Bouchard, K., Napoleão, B.: Automatic user story generation: a comprehensive systematic literature review. International Journal of Data Science and Analytics  (06 2024). \doi{10.1007/s41060-024-00567-0}

\bibitem{shahmirzadi2018textsimilarityvectorspace}
Shahmirzadi, O., Lugowski, A., Younge, K.: Text similarity in vector space models: A comparative study (2018), \url{https://arxiv.org/abs/1810.00664}

\bibitem{siddique2022ai_requirements}
Siddique, I.M.: Harnessing artificial intelligence for systems engineering: Promises and pitfalls. European Journal of Advances in Engineering and Technology  \textbf{9}(9),  67--72 (2022). \doi{https://doi.org/10.5281/zenodo.11545453}

\bibitem{Vayyavur2024_25}
Vayyavur, R.: Why ai projects fail: The importance of strategic alignment and systematic prioritization. International Journal of Research (IJR)  \textbf{11}(08),  386--391 (August 2024). \doi{10.5281/zenodo.13370566}, \url{https://doi.org/10.5281/zenodo.13370566}, received: 2 August 2024; Revised: 13 August 2024; Accepted: 22 August 2024

\bibitem{wang_review_2023}
Wang, J., Chen, Y.: A {Review} on {Code} {Generation} with {LLMs}: {Application} and {Evaluation}. In: 2023 {IEEE} {International} {Conference} on {Medical} {Artificial} {Intelligence} ({MedAI}). pp. 284--289 (Nov 2023). \doi{10.1109/MedAI59581.2023.00044}, \url{https://ieeexplore.ieee.org/abstract/document/10403378}

\end{thebibliography}
\end{document}